\documentclass[conference]{IEEEtran}
\IEEEoverridecommandlockouts
\usepackage{cite}
\usepackage{amsmath,amssymb,amsfonts}
\usepackage{algorithmic}
\usepackage{graphicx}
\usepackage{textcomp}
\usepackage{xcolor}
\usepackage{float}
\usepackage{stfloats}
\usepackage{hyperref}
\ifCLASSINFOpdf
\else
\fi

\begin{document}

\title{ Direction of Arrival estimation with Virtual Antenna Array using FMCW Radar Simulated Data }

\author{

\IEEEauthorblockN{Emre Kurto\u{g}lu and M. Mahbubur Rahman} 

\IEEEauthorblockA{
\textit{Laboratory of Computational Intelligence for RADAR (CI4R)}\\
\textit{Dept. of Electrical and Computer Engineering, The University of Alabama, Tuscaloosa, USA} }
}

\maketitle

\begin{abstract}
The FMCW radars are widely used for automotive radar systems. The basic idea for FMCW radars is to generate a linear frequency ramp as transmit signal. The difference frequency, (i.e., beat frequency) between the transmitted and received signal is determined after down conversion. The FFT operation on beat frequency signal can recognize targets at different range and velocity. Increasing demand on safety functionality leads to the Direction of Arrival (DOA) estimation to resolve two closely located targets. Consequently, the problem of angle estimation for 77GHz FMCW automotive radar simulated data has been investigated in this term project. In particular, we examined the performances of FFT, MUSIC and compressed sensing in angle estimation task, and it was found that although FFT is the fastest algorithm, it has very poor angular resolution when compared with others which are both super resolution algorithms. The code for this project report is available at \href{https://github.com/ekurtgl/FMCW-MIMO-Radar-Simulation}{https://github.com/ekurtgl/FMCW-MIMO-Radar-Simulation}.
\end{abstract}

\IEEEpeerreviewmaketitle

\section{Introduction}
Millimeter-wave technology has found great applicability in automotive radar. The typical frequency band of
millimeter-wave automotive radar is 76-81 GHz. The high frequencies allow for small enough antennas that can
fit behind the bumper of the vehicle. Also, the wide available bandwidth enables high target range resolution \cite{ref1}.
State-of-the-art automotive radar transmits FMCW at millimeter-wave frequencies, which allows for high resolution
target range and velocity estimation at much lower cost than Light Detection and Ranging (LiDAR) technology.
Automotive radar for autonomous driving needs to have high angle discrimination capability. Employing a large
antenna array would improve angular resolution, however, the resulting large package size would make integration
on the vehicle difficult. While for conventional phased-array radar a small package size implies low angular
resolution, for MIMO radar \cite{ref2}, package size is not a limiting factor. This is because MIMO radar can synthesize
virtual arrays with large aperture using only a small number of transmit and receive antennas \cite{ref3}.
Automotive MIMO radar uses FMCW waveforms along with some mechanism that guarantees waveform orthogonality. 

Through this term project, we have designed an FMCW  radar simulator that operates on 77GHz with 150 MHz bandwidth. Two distinct moving target with two different constant velocities were recognized using cell-averaging CFAR on range-Doppler map. Then three different methods namely, FFT based,  multiple signal classification (MUSIC) and Compressed sensing based algorithm have been utilized for DOA estimation. 

In section II, the details of the FMCW radar simulations are presented. Section III describes the CA-CFAR detection whereas  FFT, MUSIC and Compressed sensing based DOA estimation are presneted in section IV. Finally, the paper is finished in section V with a conclusion.

\section{FMCW Radar Simulation}

\begin{figure}[t!]
\centering
\includegraphics[width=6.7cm]{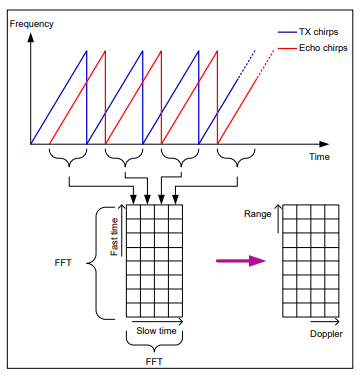}
  \caption{FMCW radar chirps. The range and Doppler estimation is performed using a 2-D FFT.}
  \label{fig:fmcw}
\end{figure}

An FMCW waveform, also referred to as chirp, is a complex sinusoid whose frequency increases linearly with time
$t  \epsilon  [0, T]$, i.e.,
\begin{equation}
\label{ft}
 f_T (t) = f_c + \frac{B}{T}t
 \end{equation}
where B is the signal bandwidth and fc is the carrier frequency. In this simulation, 77 GHz carrier frequency and 150 MHz Bandwidth were used. FMCW radar transmit chirps in a periodic fashion, with a period referred to as pulse repetition interval (PRI). The frequency of an FMCW signal over multiple periods, with PRI equal to $T$ (10 $\mu$ sec.), is shown in Fig \ref{fig:fmcw}.  One transmitter and 8 uniformly spaced linear phase array receivers were considered in this simulation. The distance between transmitter and the first receiver was 2$\lambda$ and between two consecutive receivers were $\lambda$/{2}. A total of 10 coherent processing intervals (CPI) were considered and in each CPI, a total of 256 chirps were transmitted with each pulse having 256 ADC sample bins. Two distinct targets with range 50 and 100 m and  radial velocity of 10 and -15 m/s were considered. The targets are in the far-field at an angle of -15 and 10 degrees with respect to the origin. The target's echo ranged at 50m and 100m at the radar
receiver contain a delayed and attenuated copy of the transmitted chirp. For a target at range R, moving with
radial speed of $v$, the delay equals \begin{equation}
\label{Velocity}
 \tau = \frac{2(R+vt)}{c}   
\end{equation}
where time t spans multiple periods and c is the speed of light. For the target at 50 m moving with a velocity of 10 m/s, this delay is about $0.3$ $\mu$ sec and for for the furthest target at 100 m moving with a velocity of 15 m/s, the dealy is about $0.66$ $\mu$ sec.  

\begin{figure}[t!]
\centering
\includegraphics[width=8.7cm]{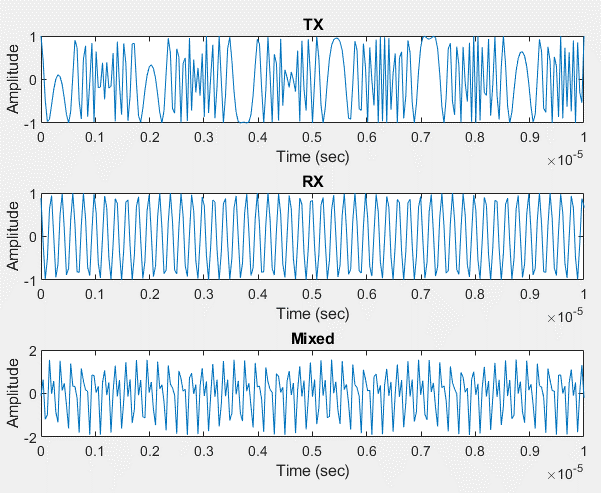}
  \caption{Transmitted, Received and mixed signals.}
  \label{fig:tx-rx}
\end{figure}

\begin{figure}[t!]
\centering
\includegraphics[width=8.7cm]{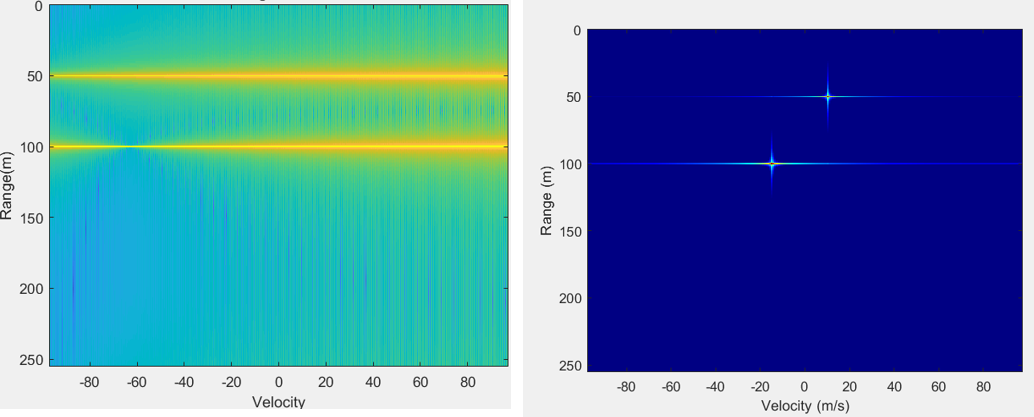}
  \caption{Range Profile (left) and Range-Doppler map (right).}
  \label{fig:range_dopp}
\end{figure}

The received signal is mixed with the transmitted chirp, which results in a complex sinusoid, known as the beat
signal. The beat signal frequency equals $ f_b = f_R + f_D $, where $f_R$ = $2RB$/ $T c$ is the range frequency and $f_D$ = $\frac{2v}{c}$$f_c$  
is the Doppler frequency. The process of obtaining the beat signal is implemented in the radio-frequency (RF)
domain by a mixer followed by a band pass filter (BPF) with maximum cut off frequency $f_b ^{max}$
, the latter filter
is used to remove signals with frequencies outside the band of interest, which also places a limit to the maximum
detectable range. The received signal from the two targets were added together to get the combined complex signals as shown in figure \ref{fig:tx-rx}. We sample the beat signal and put the samples of each chirp in the columns of a matrix, then the
row indices of that matrix correspond to fast time and the column indices to slow time (see Fig. 1). By applying
FFTs on the sampled beat signal along fast time, one can identify $f_R$, based on which the target’s range can be
obtained as $R$ = $cf_RT$ /$(2B)$. To obtain the target’s Doppler frequency, a second FFT operation is subsequently
carried out along the slow time (the range frequency fR is the same across slow time). The application of these
two FFTs is equivalent to a two-dimensional (2-D) FFT of the beat signal in fast and slow time, and the result is
called range-Doppler spectrum as depicted in figure \ref{fig:range_dopp}. Range and Doppler detection can be performed using conventional thresholding
based methods applied to 2-D range-Doppler spectrum, such as the constant false alarm rate (CFAR) detector \cite{ref4},
or the recently proposed deep neural network based detector \cite{ref5}. Via the 2-D FFT, the targets can be separated
in the range and Doppler domain. Since the number of targets within the same range-Doppler bin is small, angle
finding can be carried out using sparse sensing techniques, such as compressive sensing.

\section{Cell Averaging Constant False Alarm Rate (CA-CFAR) Detection}

CA-CFAR is one of the most popular CFAR detectors. In most cases, it is used as a baseline comparison method for other CFAR techniques. In this work, we apply CA-CFAR to RD maps to detect true target range and velocity. In the CA-CFAR detector, the decision for a given cell (also called a \textit{cell under test} (CUT)) is given based on comparison with a threshold, $T$, given by
\begin{equation}
\label{threshold}
T=aP_n
\end{equation}
where $a$ is a scaling factor (also called the threshold factor) and $P_n$ is the noise power estimate.  The noise power is estimated from leading and lagging neighboring cells. 

\begin{figure}[b!]
\centering
\includegraphics[width=8.7cm]{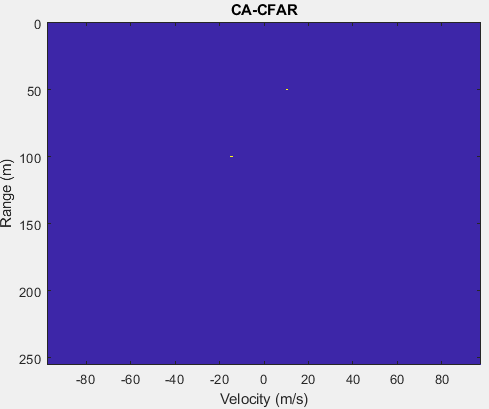}
  \caption{CA-CFAR detection for targets located at 50 and 100 meters with radial velocities of 10 and -15 m/s respectively.}
  \label{fig:cfar}
\end{figure}

From Eq. (\ref{threshold}), it may be observed that the adaptive threshold varies according to the data. Moreover, the desired probability of false alarm ($P_{fa}$) rate can be kept at a constant level using a proper threshold factor, $a$. The noise power estimate can be computed as

\begin{equation}
\label{noise}
P_n = \frac{1}{M} \sum_{m=1}^{M} x_{m} 
\end{equation}
where $M$ is the number of training cells and $x_m$ is the sample in each training cell. The threshold, $a$, then can be written as
\begin{equation}
\label{cfar_th}
a = M(P_{fa}^{-1/M}-1)
\end{equation}
In total, we used 56 training and 25 guard cells for detection and the adaptive threshold was able to detect ranges and velocities of the targets successfully as can be seen in Figure \ref{fig:cfar}.

\section{Angle Estimation}
\begin{figure}[b!]
\centering
\includegraphics[width=8.7cm]{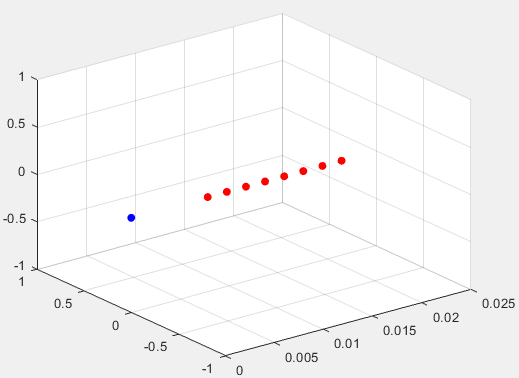}
  \caption{TX and RX antenna locations.}
  \label{fig:txrx}
\end{figure}

Once we know where the targets are located in a RD map, we can perform angle estimation with an increased SNR. In this work, we did all of our analysis using 1 transmitter (TX) and 8 receivers (RX) with $d = \lambda/2$ spacing where $\lambda$ is the wavelength. Locations of TX and RX antennas in 3D coordinates can be found in Figure \ref{fig:txrx}.

\subsection{Using FFT}
\begin{figure}[t!]
\centering
\includegraphics[width=8.7cm]{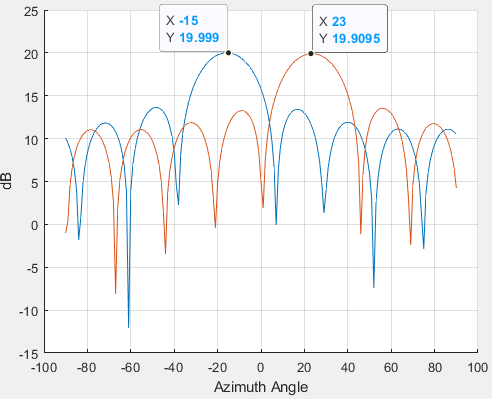}
  \caption{Estimated target angles by using FFT.}
  \label{fig:fftspect}
\end{figure}

Fast Fourier transform (FFT) is an algorithm that computes the discrete Fourier transform (DFT) of a sequence in an efficient way. Since it is a computationally efficient method, it can be employed in many near-real-time signal processing applications. After computing RD maps and CFAR detected points, we can perform FFT across channels to retrieve the angle information. In this work, we placed two targets at 10 and -15 degrees and the estimated target angles can be found in Figure \ref{fig:fftspect}. It can be observed that although the estimated angle for the second target is correct, it is not the case for the first target which shows peak at 23 degrees. 

\subsection{Using MUSIC}
Although FFT is an efficient and easy to implement method, in most cases, its resolution is not sufficient to separate closely located targets. MUltiple SIgnal Classification (MUSIC) is a high-resolution direction-finding algorithm based on the eigenvalue decomposition of the sensor covariance matrix observed at an array. MUSIC belongs to the family of subspace-based direction-finding algorithms. Assume that there are $D$ uncorrelated or partially correlated signal sources. The sensor data consists of the signals, as received at the array, together with noise. A sensor data snapshot is the sensor data vector received at the $M$ elements of an array at a single time $t$. The received signal can be written as

\begin{equation}
x(t) = As(t) + n(t)
\end{equation}
where
\begin{equation}
s(t) = [s_1(t),s_2(t),...,s_D(t)]' 
\end{equation}
\begin{equation}
A = [a(\theta_1),a(\theta_2),...,a(\theta_D)]
\end{equation}
Here, $x$ is an M-by-1 vector of received snapshot of sensor data which consist of signals and additive noise, $A$ is an $M$-by-$D$ matrix containing the arrival vectors. An arrival vector consists of the relative phase shifts at the array elements of the plane wave from one source. Each column of $A$ represents the arrival vector from one of the sources and depends on the direction of arrival, $\theta_D$. $\theta_D$ is the direction of arrival angle for the $d^{th}$ source and can represents either the broadside angle for linear arrays or the azimuth and elevation angle for planar or 3D arrays, $s(t)$ is a $D$-by-1 vector of source signal values from $D$ sources and $n(t)$ is an $M$-by-1 vector of sensor noise values. The covariance matrix, $R_x$, can then be derived from the received signal data. When the signals are uncorrelated with the noise, the sensor covariance matrix has two components, namely the signal covariance matrix and the noise covariance matrix. The sample sensor covariance matrix is an average of multiple snapshots of the sensor data. 

\begin{equation}
R_x = E\{xx^H\} = \frac{1}{T} \sum_{k=1}^{T} x(t)x(t)^H
\end{equation}
where T is the number of snapshots. Once we obtain the covariance matrix, we can do eigenvalue decomposition to obtain noise and signal subspaces, $U_n$ and $U_s$. MUSIC works by searching for all arrival vectors that are orthogonal to the noise subspace. To do the search, MUSIC constructs an arrival-angle-dependent power expression, called the MUSIC pseudospectrum:

\begin{equation}
P_{MUSIC}(\overrightarrow{\phi}) = \frac{1}{a^H(\overrightarrow{\phi})U_nU_n^Ha(\overrightarrow{\phi})}
\end{equation}

\begin{figure}[t!]
\centering
      \includegraphics[width=8cm]{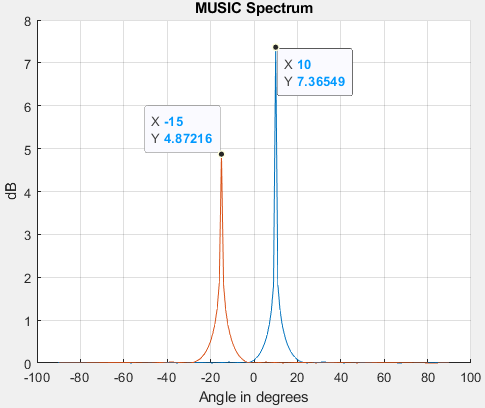}
      \caption{Estimated target angles by using MUSIC.}
      \label{fig:musicspect}
\end{figure}

\begin{figure}[t!]
\centering
      \includegraphics[width=8cm]{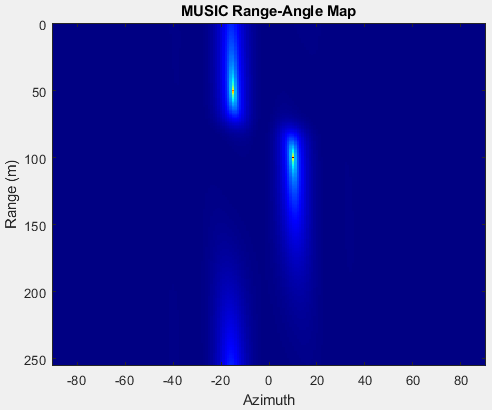}
      \caption{Range-angle map computed by MUSIC.}
      \label{fig:musicmap}
\end{figure}

When an arrival vector is orthogonal to the noise subspace, the peaks of the pseudospectrum are infinite.  In practice, because there is noise, and because the true covariance matrix is estimated by the sampled covariance matrix, the arrival vectors are never exactly orthogonal to the noise subspace. Then, the angles at which $P_{MUSIC}$ has finite peaks are the desired directions of arrival. Because the pseudospectrum can have more peaks than there are sources, the algorithm requires that you specify the number of sources, $D$, as a parameter. Then the algorithm picks the $D$ largest peaks. For a uniform linear array (ULA), the search space is a one-dimensional grid of broadside angles. For planar and 3D arrays, the search space is a two-dimensional grid of azimuth and elevation angles. Resulting pseudospectrums for two targets located at 10 and -15 degrees can be found in Figure \ref{fig:musicspect}. Range-angle maps can further be computed by first taking 1D range FFT of the radar data cube and then apply MUSIC for each range bin. One sample of range-angle maps can be found in Figure \ref{fig:musicmap}.

\subsection{Using Compressed Sensing}
Compressed sensing (CS) has recently gained enormous popularity in lots of different fields. It can accurately
recover the signal with only a few random samples, provided that the signal is sparse or compressible in a certain domain. In this work, we aim to make use of the CS technique to estimate angles of the targets. Considering
the fact that in radar, the range and Doppler processing are usually performed before target DOA estimation and the
number of targets fallen in the same range-Doppler cell are limited, the CS assumption of sparsity is hence satisfied, which makes the application of CS feasible. 

Let $s$ denote the original unknown signal with dimension
of $N$×1. Now $s$ experiences compressive sensing and we
get 
\begin{equation}
y = \Phi s
\label{cs}
\end{equation}
where $\Phi$ is \textit{M}x\textit{N} (\textit{M}$\ll$\textit{N}) sensing matrix. The
signal \textit{y} represents the obtained under-sampled data with dimension \textit{M}x1.  

\begin{figure}[t!]
\centering
      \includegraphics[width=8cm]{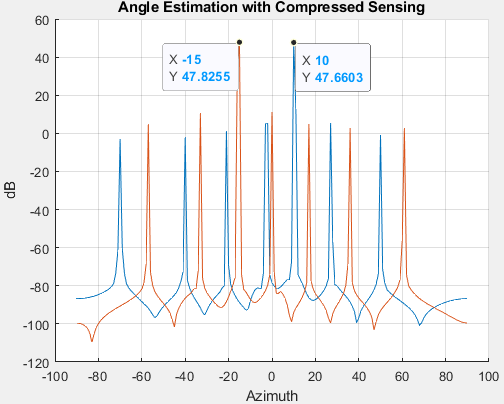}
      \caption{Estimated target angles by using CS.}
      \label{fig:csspect}
\end{figure}

In \ref{cs}, if $s$ is sparse or compressible in a certain domain and has only $K$ (\textit{K}$\ll$\textit{N}) non-zero entries in that domain, then we say $s$ is \textit{K}-sparse. CS theory shows that the original \textit{K}-sparse signal $s$ can be precisely reconstructed from the severely incomplete measurement $y$.

L1 norm convex optimization algorithm is a candidate for the estimation of $s$. Because $s$ is sparse, it can be estimated by the following optimization problem

\begin{equation*}
\begin{aligned}
& \min_{s} ||s||_1 \\
& \text{subject to}
& & ||y-\Phi s||_2 \leq 1 
\label{min}
\end{aligned}
\end{equation*}

CVX toolbox \cite{gb08} is employed to perform L1 minimization. Resulting angle estimation plot for CS can be found in Figure \ref{fig:csspect}.

\section{Conclusion}
This paper investigated the problem of DOA estimation for automotive FMCW radar to resolve two targets at close proximity. Two moving targets were simulated and the return signals were received with 8 uniformly spaced linear phase array antenna. Consequently, 8 channels simulated data were generated by applying the  principal of FMCW radar. While CA-CFAR has been used to identify moving targets, FFT, MUSIC and CS were used to estimate the angles of arrivals. It is observed that FFT is the fastest algorithm amongst them but it also has the poorest angular resolution which makes it unattractive for the applications that require high resolution. It can be useful for applications where high resolution is not needed and speed has priority (e.g. real-time applications). MUSIC and CS, on the other hand, are both super resolution algorithms and they are able to separate closely located targets. Their biggest disadvantage is of course is the computational cost. While MUSIC needs to to scan through the angle grid, CS has to do iterations to minimize the L1 norm which increases the computational burden.

\section*{Acknowledgment}
The authors would like to thank Dr. Shunqiao Sun
for his immense support during the completion of this project.

\bibliographystyle{IEEEtran}
\bibliography{ref}

\end{document}